\documentclass[pre,preprint,elide,hyperlink,sort&compress]{revtex4}
\pdfoutput=1
\usepackage[varg]{txfonts}
\usepackage{graphicx}
\usepackage[usenames]{color}
\usepackage{hyperref}

\newcommand{\parref}[1]{(\ref{#1})}

\newcommand{\vect}[1]{\mathbf{#1}}
\newcommand{\units}[1]{\,\mathrm{#1}}
\newcommand{\dexp}[1]{\cdot 10^{#1}}
\newcommand{\neigh}{\mathrm{neigh}}
\newcommand{\branch}{\mathrm{branch}}
\definecolor{DarkRed}{rgb}{0.6,0,0}
\definecolor{Blue}{rgb}{0.0,0.0,0.75}
\definecolor{Red}{rgb}{0.75,0.0,0.0}
\definecolor{Violet}{rgb}{0.75,0.0,0.75}
\definecolor{Green}{rgb}{0.0,0.5,0.0}

\newcommand{\mysubsubsection}[1]{\textit{#1}}
\begin{document}

\title
[Discharge trees with self-consistent charge transport]
{Growing discharge trees with self-consistent charge transport: the collective dynamics of streamers}
\author{Alejandro Luque$^1$ and Ute Ebert$^{2,3}$}
\affiliation{$^1$~Instituto de Astrof\'isica de Andaluc\'ia (IAA), CSIC, Granada, Spain,\\
$^2$~CWI, P.O. Box 94079, 1090 GB Amsterdam, The Netherlands,\\
$^3$~Department of Physics, Eindhoven University of Technology,
  The Netherlands.}

\begin{abstract}

%
We introduce the generic structure of a growth model for branched discharge trees
that consistently combines a finite channel conductivity with the physical law
of charge conservation.
It is applicable, e.g., to streamer coronas near tip or wire electrodes
and ahead of lightning leaders, to leaders themselves and to the complex
breakdown structures
of sprite discharges high above thunderclouds. Then we implement and
solve the simplest model
for positive streamers in ambient air with self-consistent charge transport.
We demonstrate that charge conservation contradicts the common assumption of dielectric breakdown models that the electric fields inside all streamers are equal
to the so-called stability field and we even find cases of 
local field inversion.  We also discuss the charge distribution inside discharge
trees, which provides a natural explanation for the observed reconnections
of streamers in laboratory experiments and in sprites.
Our simulations show the structure of an overall ``streamer of streamers'' that we name collective streamer front, and predict
effective streamer branching angles, the charge structure within streamer trees,
and streamer reconnection.
\end{abstract}

\maketitle


\section{Introduction}

\subsection{Phenomena and state of understanding}

When a high electric voltage is suddenly applied to ionizable matter,
electric breakdown frequently takes the form of growing filaments,
and these filaments can form a complex tree structure. Discharge trees are
observed in streamer coronas around tip or wire electrodes, in the streamer
coronas ahead of propagating lightning leaders \cite{Rakov2003/ligh.book}
and in the (hot) leaders themselves.
Streamer discharge trees also appear in transient luminous events
such as jets \cite{Wescott1995/GeoRL}, gigantic jets \cite{Su2003/Natur}
and sprites \cite{Franz1990/Sci}
between thunderclouds and the
ionosphere.
Streamer and leader trees are a generic response to high voltage pulses;
they appear in various gases, liquids and solids in plasma and high voltage
technology.

Our understanding of such non-thermal, filamentary electrical discharges is remarkably unbalanced.  On
the one hand, we are now reaching a very detailed knowledge on their
microphysics; this includes
models of electron energy distributions \cite{Li2012/JCoPh}, and of
transport coefficients and cross-sections of the main reactions,
at least for air and other common gas compositions.  This
knowledge translates into sophisticated and reasonably
accurate models of single streamers
\cite{Eichwald2008/JPhD,Pancheshnyi2005/PhRvE,Qin2012/GeoRL,Liu2010/GeoRL,
Luque2012/JCoPh,Li2012/PSST}, the initiation of streamer branching
\cite{Li2012/JCoPh,Luque2011/PhRvE,Li2012/PSST} and the merging of two nearby
streamers \cite{Luque2008/PhRvL,Bonaventura2012/PSST}.
On the other hand, we barely understand most macroscopic processes in a fully
developed corona or streamer tree involving hundreds or thousands of
mutually interacting plasma filaments.  The large scale transport of
charge, the internal electric fields and the influence of the many surrounding
streamers on one single streamer are rarely discussed in
the literature.  However, these mechanisms are relevant for
the propagation of long sparks
\cite{Raizer1991/book,Bondiou1994/JPhD,Kochkin2012/JPhD} and
the approach of lightning leaders towards protecting rods.
The overall tree structure also determines which volume fraction
of the medium is ``treated'' by the discharge, creating radicals, ions
and subsequent chemical products relevant for plasma technology and
for the production of greenhouse gases during a thunderstorm.

Most studies on the growth of electrical discharge trees descend from
the Dielectric Breakdown Model (DBM) \cite{Niemeyer1984/PhRvL}
that Niemeyer \textit{et al.}\ proposed in 1984 to explain
the fractal properties of some electrical discharges such as
Lichtenberg figures that propagate over a dielectric surface.  In their model, a discharge tree expands in discrete time-steps
by the stochastic addition of new segments with a probability that depends on the local electric field.

We are not aware of many models of fully three-dimensional streamer trees not based on the DBM.  Only Akyuz \textit{et al.} \cite{Akyuz2003/JElec} modeled streamers as a tree of connected, perfectly conducting cylinders that propagate according to simple rules based on the value of the electric field surrounding
the tips. The computations required to solve the electrostatic problem
limited their simulations to small trees with less than 10 branches.

The original DBM as well as \cite{Akyuz2003/JElec} assume that the channels in the tree are perfectly conducting, but there is strong experimental evidence that the electric potential decreases along a discharge channel.

\subsection{Electric fields inside discharge trees: stability field
versus self-consistent charge transport}

The common approach to introduce a potential decay along a streamer channel and inside the streamer corona is to assume that the electric field inside a streamer has a fixed value, the so-called stability field. E.g., in air at standard temperature and pressure the stability field of positive streamers is thought to be 4 to 5~kV/cm. A fixed stability field is used to model the streamer corona that precedes a leader in a long spark discharge~\cite{Goelian1997/JPhD,Becerra2006/JPhD/1,Arevalo2011/JPhD}
or the enormous streamer trees in sprite discharges high above thunderstorms~\cite{Pasko2000/GeoRL}.


However, the concept of a fixed field inside streamer channels lacks any theoretical support.  Rather, it is based on a phenomenological interpretation of experiments that nevertheless have not measured the internal streamer fields.
  Originally, the concept of stability field refered to the minimum average applied field for sustained streamer propagation in a gap between parallel electrodes \cite{Allen1995/JPhD}.  The existence of such a minimum field around
4 to 5~kV/cm was interpreted \cite{Gallimberti1972/JPhD,Gallimberti1979/JPhys} in terms of a now discarded model of streamers as isolated patches of charge.  Later it was found that the relation between the applied potential at the originating electrode $U$ and the longest streamer length $L$ is roughly linear with $U/L \approx (4.5-5)\units{kV/cm}$ in air \cite{Bazelyan2010/book}.  Since this value was close to the existing concept of an stability field, the results were interpreted as indicating that the stability field was the electric field inside the streamer channel.  However, even the earliest numerical simulations of 2d streamers \cite{Dhali1987/JAP} already showed a clearly non-constant electric field in the channel.  As we will see, this variation is enhanced by the collective dynamics of a streamer tree.  Indeed, 
our results will show that the assumption of a constant electric field in all streamers is in contradiction with a consistent charge transport model, as long as conductivity stays finite.

Recent simulations of density models resolving the inner structure of streamers already have established the relevance of a self-consistent charge transport model for the dynamics of streamer
channels, and, in particular, for the dynamics of the electric field in the channel.  For upper-atmospheric streamers, Liu \cite{Liu2010/GeoRL} and Luque and Ebert \cite{Luque2010/GeoRL} independently showed that the re-brightening of sprite
streamer trails is due to a second wave associated
with a significant increase of the electric field in the
sprite channel; Luque and Gordillo-V\'azquez \cite{Luque2011/GeoRL}
postulated later that sprite beads are also caused by persisting and
localized electric fields.  These
electric fields may only persist due to a finite conductivity in the
streamer channel \cite{Gordillo-Vazquez2010/GeoRL}, which also
sets their decay times.

To our knowledge, the only DBM-inspired models that treat
the charge transport self-consistently appear in the context of
discharge trees in dielectrics \cite{Noskov2001/JPhD}, generated when
a solid insulator is subjected to an intense, repetitive electrical stress \cite{Dissado1993/PhRvB}.

\subsection{Content of the paper}

In the present paper we first outline the general structure of a model
for growing discharge trees that consistently incorporates charge conservation.
Then we introduce the simplest model for a streamer corona as a tree structure
of linear channel segments
with a finite fixed diameter and with a finite fixed conductivity.  The
streamer channel tips advance and branch according to simple,
phenomenologically motivated rules.  We analyze
the internal electric fields and the transport of charge in fully
branched, extensive streamer coronas.  This is a
stepping stone towards more realistic and detailed models and, although
many improvements of our approach are straightforward, we have often
kept complexity at a minimum in order to focus on the
overall qualitative behavior of streamer trees with realistic
conductivities and consistent charge transport, which appears
to be largely unexplored in the existing literature.

The paper is organized as follows: in section \ref{sec:model} we give general prescriptions for discharge tree models with self-consistent charge transport, which are then particularized into the simplest streamer tree model, which we have implemented.  We present the most relevant results of the model in section \ref{sec:results}.  Finally, section \ref{sec:summary} concludes with a short summary and discussion.

\section{Description of the model}
\label{sec:model}

\subsection{The structure of a growing tree model that conserves electric charge}

We model the discharge tree as a growing network of conductors,
with an emphasis on charge conservation and transport within the tree.
The geometric structure of the network with its charge content and
the external electric field determine the actual electric field distribution;
this field distribution together with the conductivity distribution
within the network determine the consecutive charge transport in the tree,
and the local field distribution at the tip determines growth and branching
of the tree tips. The tip dynamics determines diameter, conductivity and
tree structure of the newly grown parts of the network.

Let us now discuss the general structure of such a model with reasonable
approximations, before introducing the simplest manifestation of such a model
in the next subsection.

\begin{figure}
\includegraphics[width=7cm]{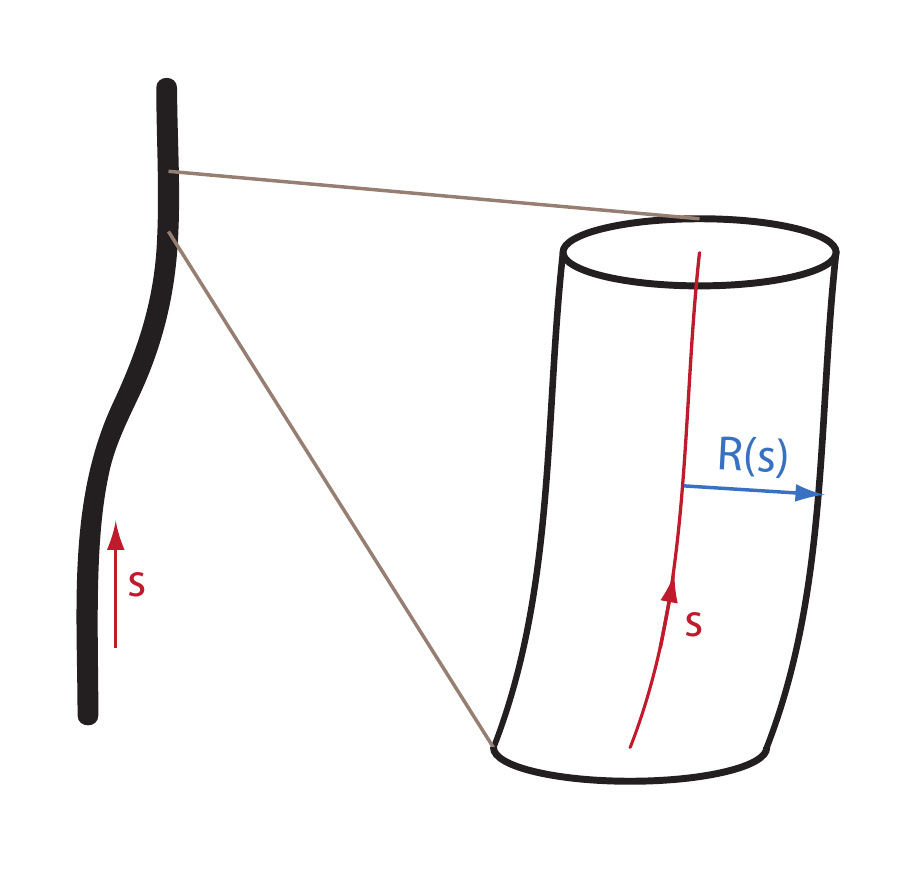}
\caption{\label{channel}  Schematic of a part of a
discharge channel, parameterized by arc length $s$ and radius $R(s)$.
The interior of the channel is filled by a mostly electrically neutral
plasma providing the conductivity of the channel while the lateral walls
contain most of the electric charge that is due to an overshoot of
plasma species of one polarity.}
\end{figure}

\mysubsubsection{Linear channel parts: radius $R$, line charge density $q$,
line conductivity $\sigma$ and electric current $I$. ---}
A schematic of a linear channel part is provided in Fig.~\ref{channel}.
We parameterize the channel length with a longitudinal or arc length coordinate $s$,
and we assume these parts to be cylindrically symmetric with a radius $R(s,t)$.
The conductivity of the channel is provided by the densities $n_{e,\pm}$ and
the mobilities $\mu_{e,\pm}$ of the electrons and of the positive and negative ions
inside the channel, and we define a line conductivity $\sigma(s,t)$ as the integral
of the conductivity over the channel cross section
\begin{equation} \label{linesigma}
\sigma(s,t) = \int 2\pi\;r\;dr\;\left(\mu_en_e + \mu_+n_+ + \mu_-n_-\right)(r,s,t),
\end{equation}
and a line charge density $q(s,t)$ by the integral of the charge density over the channel cross section
\begin{equation} \label{linecharge}
q(s,t) = \int 2\pi\;r\;dr\;{\rm e}\;\left(n_+ - n_e - n_-\right)(r,s,t),
\end{equation}
where e is the elementary charge. The line conductivity is the inverse of the resistance per length, and the line charge density is the charge per length.

In general, we can assume that the radius varies slowly over the arc length $s$.
The electric charge typically resides in the surface of channel. It can be assumed
to be cylindrically symmetric as long as other charges stay at a distance much larger
than the channel radius. According to standard electrodynamics, the electric field
created by the charge of the channel is determined only by the line charge density
and not by the channel radius at distances much larger than the channel radius.

The electric current $I(s,t)$ along the conducting channel is determined by Ohm's law,
\begin{equation}
I(s,t) = \sigma(s,t)\;E(s,t),
\end{equation}
where $E(s,t)$ is the local electric field inside the channel;
here we used that the electric field inside the channel, i.e.,
inside the space charge layer, does essentially not change
in the radial direction, and that it is oriented along
the channel~\cite{Ratushna/prep} --- otherwise the current would flow
into or out of the channel walls and would change the charge content
very rapidly; hence as long as charges change slowly, the field is
directed along the axis.

The conservation of electrical charge implies
\begin{equation}
\partial_t q(s,t) + \partial_s I(s,t)=0.
\end{equation}

For radius $R(s,t)$ or line conductivity $\sigma(s,t)$ particular
dynamical equations could be implemented that incorporate physical
understanding of the channel dynamics. Alternatively they can be
considered as fixed after they have been generated by the motion
of the channel head.

\mysubsubsection{Head radius, charge, velocity and branching. ---}
The charge distribution in the discharge head and channel together
with the external field determine the electric field distribution
at the head. The head velocity in general depends not only on
the electric field in some particular spot, but on the electric field
$\vect{E}_{enh}$ and electron density distribution in the whole ionization 
region
at the discharge head; and the shape of this region is strongly determined
by the head radius $R$. The velocity of head or tip can therefore
be considered as a function of radius $R$, electric field $\vect{E}_{enh}$,
polarity $\pm$ and of gas type and conditions,
\begin{equation}
{\bf v}_{\rm tip}^\pm = {\bf v}^\pm\Big(\vect{E}_{enh},~R,
\mbox{ gas type and preionization}\Big).
\end{equation}
For the velocity of streamers in air, Naidis has suggested a particular
analytic approximation in~\cite{Naidis2009/PhRvE}.

For branching of the channel tip, an appropriate distribution
as a function of the head parameters has to be found. For positive
streamers in air, both
experimental~\cite{Briels2008/JPhD/1,Nijdam2008/ApPhL,Heijmans2013/arXiv,Kanmae2012/JPhD}
and theoretical~\cite{Luque2011/PhRvE} studies have been presented;
they constitute the start of quantitative investigations.

The channel conductivity is also created at the channel tip.
Particular results for ionization degrees for streamers in air
will be discussed later. For leaders, also a reduced medium density due to
thermal expansion contributes to increasing the electrical conductivity
of the channel.

\mysubsubsection{Electric field. ---} The electric field is given by
the external field plus contributions due to the charges in the tree.
In density approximation, the electric potential is given by the classical equation
\begin{equation}\label{E-density}
\phi({\bf r}) = \phi_{ext}({\bf r}) + \frac1{4\pi\epsilon_0}\int d{\bf r}'\;
\frac{{\rm e}(n_+-n_e-n_-)({\bf r}')}{|{\bf r}-{\bf r}'|}.
\end{equation}
We recall that the electrical charge density ${\rm e}(n_+-n_e-n_-)$ is
nonvanishing essentially only in the walls of the channels, at the radius $R$.
When approximating the channel by a line as above, the kernel in \parref{E-density} has to be modified by a regularization to avoid unphysical singularities for $|{\bf r}-{\bf r}'|\to0$.  We use
\begin{equation}\label{E-line}
\phi({\bf r}) = \phi_{ext}({\bf r}) + \frac1{4\pi\epsilon_0}\int ds\;
\frac{q(s)}{|{\bf r}-{\bf r}(s)|+R}.
\end{equation}
Other kernels may be acceptable as long as they have the correct asymptotics for $|{\bf r}-{\bf r}(s)| \gg R$ but we have found
problems of instability with non-monotonic kernels.

\mysubsubsection{The general set-up of this model}
allows the implementation of approximations derived
from more microscopic 3D fluid or particle models on propagation and branching
of channel heads of positive or negative polarity and on the diameters and
dynamically changing conductivities of the discharge channels. In this manner,
the model eventually can serve as an upscaling step in a hierarchy of
multiscale models for streamers, leaders, sprites, jets or any other discharge types,
into which the detailed knowledge on diameters, velocities,
ionization and branching rates derived on a smaller length scale can be implemented.
Here we recall that, e.g., for streamers, the diameters, velocities and
ionization degrees can vary by several orders of magnitude~\cite{Briels2008/JPhD/1}.


\subsection{The simplest streamer tree model}

In the current paper, we will make a number of assumptions to make
the model as simple as possible. This will allow us to identify
the key new features induced by consistent charge transport,
without having to wonder whether properties are due to particular
other model features.

In this simplest model, we assume that all channel parts and tips
have the same time independent radius $R$ and line conductivity $\sigma$,
that the streamer head velocity is proportional to the local electric field,
and that branching is a Poisson process depending on the length
of the streamer segment.

Together with the electric potential being fixed at the boundary
of the simulation domain, and with the location of the electrode
that supplies the electric current,
these assumptions characterize the physical model.

\begin{figure*}
\includegraphics[width=\linewidth]{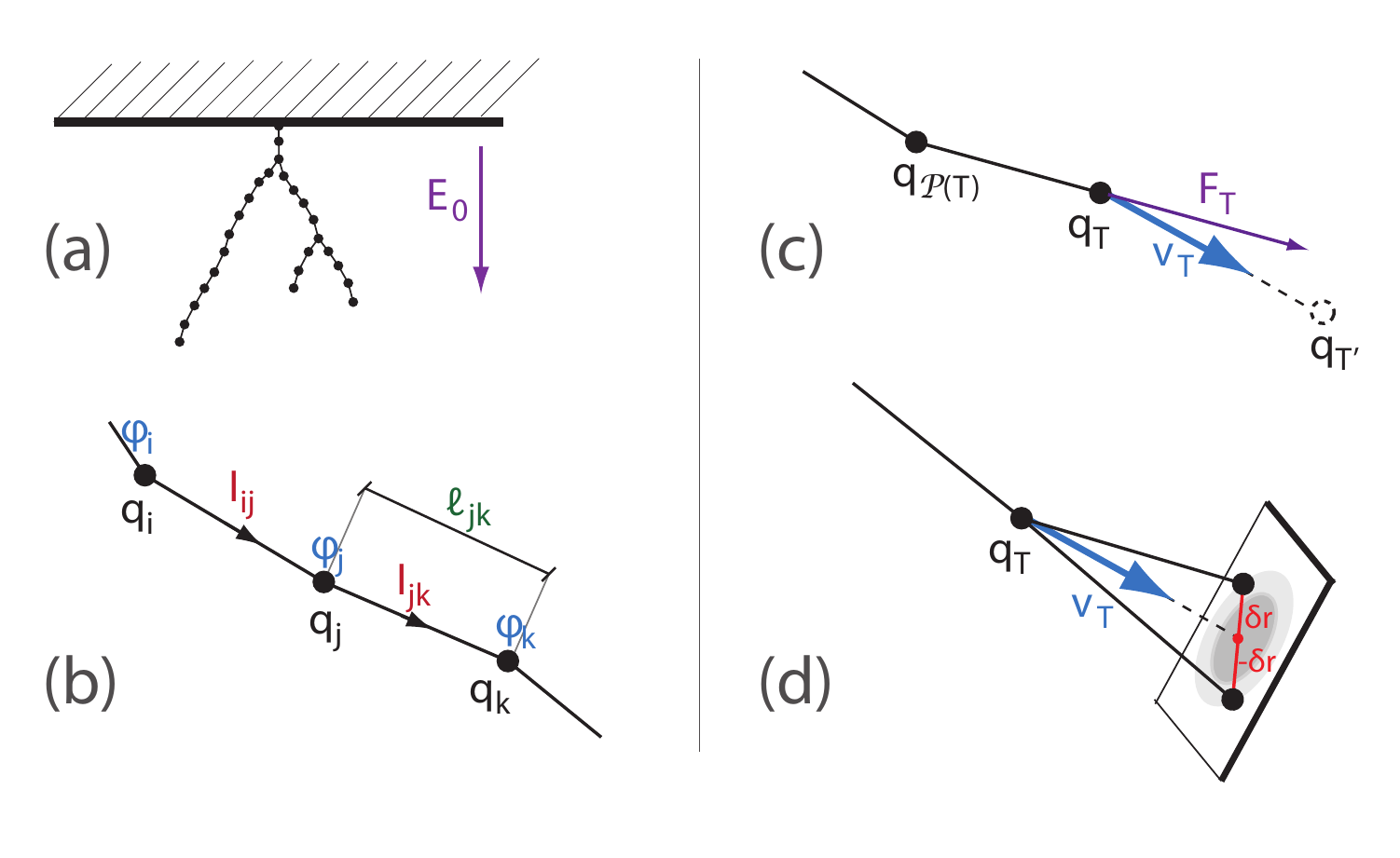}
\caption{\label{chain}  Scheme of the
   numerical implementation of the model. (a) The
  streamer tree is represented as a tree of nodes, each containing
  some charge and connected to neighbouring nodes with a
  finite-conductance link.  (b) Each node $i$ contains a charge
  $q_i$; during the relaxation phase of the numerical simulation,
  charge is transported along the conductor links of length $\ell_{ij}$
  by currents
  $I_{ij}$, changing the electric potentials $\phi_i$.
  (c) The terminal nodes of the tree advance in discrete
  time steps by the addition of a node further along the channel; the
  location of this node $T'$ is determined by a velocity $\vect{v}_T$ determined
  by the local electric field at the terminal node $T$. (d) When a
  streamer branches, the offspring of the node $T$ consists of two
  nodes: each one is displaced from the straight
  path by a random vector $\pm \delta\vect{r}$ in the
  plane perpendicular to the original streamer path;
  $\delta\vect{r}$ is drawn from a bi-dimensional gaussian
  probability distribution.}
\end{figure*}

\subsection{Numerical implementation}
We shall describe now the numerical implementation of the model described above.
This numerical implementation, along with all the input files used in this articles is freely available\footnote{Source code is accessible at 
\url{https://github.com/aluque/strees}.  For a short documentation, see
\url{http://aluque.github.io/strees/}}.

As sketched in Figure~\ref{chain}, we replace the continuous
arc lengths $s$ of the different linear channel parts by the
set $i=1,\ldots,N$ of $N$ charged nodes at positions $\vect{r}_i$,
each containing a time-dependent charge $q_i(t)$, and a time dependent
electric potential $\phi_i(t)$ is attributed to each node.
The tree evolves through two coupled mechanisms.  First, due to the electric field,
charge is transported along the edges.  Second, each channel grows or branches
at its tip  according to the local conditions.  In our model, we alternate
between these two evolutions: to
evolve our system from time $t$ to time $t + \Delta t$ we first calculate
the electric field and transport the charge in the tree for an
interval $\Delta t$, and then we add new nodes at the tips of
existing channels, allowing some
channels to branch eventually. The choice of the numerical time step
$\Delta t$ is discussed in \ref{AppNumerics}. We describe now the steps of the simulation.

\mysubsubsection{Electric field with boundary conditions. ---}
We assume that the stem of the discharge tree is connected to an upper
planar electrode located at $z=0$ that creates a constant background
electric field $\vect{E}_0$.
This electrode together with the set of charges $q_i$ with $i=1\ldots N$
within the discharge tree create an electric potential
\begin{eqnarray}\label{phi_j}
  \phi_j & = &\frac{1}{4\pi\epsilon_0}\sum_{i=-N}^N{\frac{q_i}{\ell_{ij} + R}}
  + \phi_{\rm ext}(\vect{r}_j), \\
  \nonumber
  \ell_{ij} & = & |{\bf r}_i-{\bf r}_j|,
  \quad \phi_{\rm ext}(\vect{r}) = - \vect{E}_0 \cdot \vect{r}
  \label{potentials}
\end{eqnarray}
at the node $j$, according to equation (\ref{phi}). Here $i=-1 \ldots -N$
parameterizes the mirror charges introduced to keep the electrode at potential zero:
for each charge $q_i$ located at ${\bf r}_i=(x, y, z)$ a mirror charge
$q_{-i}=-q_i$ is located at ${\bf r}_{-i}=(x,y,-z)$.
The node $i=0$ is taken as the root of the tree;
it is located at the origin, and it is discharged by the contact with the electrode.
Therefore $q_0=0$.

For details on the numerical solution of the electrostatic problem (\ref{phi_j}),
we refer to \ref{AppNumerics}.

\mysubsubsection{Charge transport within the tree. ---}
During the relaxation phase, electric currents flow along the
conductor links according to Ohm's law, where the current through
each link is calculated from the potential difference between its two
endpoints as
\begin{equation}
  I_{ij} = \sigma E_{ij},\qquad E_{ij} = -\frac{\phi_i - \phi_j}{\ell_{ij}}.
  \label{j}
\end{equation}
Due to these currents, the charge at node $i$ changes as
\begin{equation}
  \frac{dq_i}{dt} = \sum_{j\in{\neigh(i)}} I_{ij},
  \label{dqdt}
\end{equation}
where $\neigh(i)$ stands for the set of nodes connected to $i$.
For the root node $i=0$, $q_0=0$ is maintained because the current $I_{01}$ is exactly balanced by the current drawn from the electrode.

At each time step, we integrate the set of ordinary differential
equations and \parref{dqdt}, coupled with \parref{potentials},
from $t$ to $t + \Delta t$.  In our
implementation, we used the real-valued Variable-coefficient
Ordinary Differential Equation (VODE) solver \cite{Brown1989/JSSC}.

\mysubsubsection{Growth of tree tips. ---}
\label{sec:growth}
Each streamer in the tree grows at its tip,
and we model this growth by adding a new node $T'$ ahead of the old terminal node $T$
after time $\Delta t$ at the location
\begin{equation}
  \vect{r}_{T'} = \vect{r}_T + \vect{v}_T \Delta t,
  \label{stepping}
\end{equation}
see figure \ref{chain}c.

The tip velocity $\vect{v}_T$ depends on the electric field distribution
around the terminal node $T$. We approximate this distribution by
the electric field in the node $T$ generated by the background field and
the charges of all other nodes plus the term $\vect{F}_T$
\begin{equation}
  \vect{E}_T = \vect{E}_0 +
  \frac{1}{4\pi\epsilon_0}\sum_{j\neq T}^N{\frac{q_j \vect{e}_{jT}}
    {(|\vect{r}_j - \vect{r}_T|+ R)^2}} + \vect{F}_T,
  \label{E_T}
\end{equation}
where $\vect{e}_{jT}$ is a unit vector pointing from $\vect{r}_j$ to $\vect{r}_T$.

The term $\vect{F}_T$ accounts for the contribution of the terminal node $T$.  In the limit $\Delta t \to 0$, as the separation between nodes decreases, the charge contained in the terminal node becomes negligible compared with the many charges in the channel at distances shorter than $R$.  For finite $\Delta t$, the term $\vect{F}_T$ accounts for the contribution of these many charges that are now summed up into the terminal charge $q_T$:
\begin{equation}
   \vect{F}_T = \frac{q_T \vect{e}_{\mathcal{P}(T)T}}{4 \pi
     \epsilon_0 R^2},
   \label{F_T}
\end{equation}
where $\vect{e}_{\mathcal{P}(T)T}$ is the unit vector that points towards $T$ from its predecessor $\mathcal{P}(T)$.

We will assume that the tip velocity is proportional to
$\vect{E}_T$ through a model parameter that we name \textit{head
mobility}, $\mu_H$.  Since the charges that enter into equations
\parref{E_T} and \parref{F_T} change continuously during the time
interval $\Delta t$, we
advance the streamer tips with a velocity that is linearly
interpolated from its values at $t$ and at $t + \Delta t$:
\begin{equation}
  \label{velocity}
  \vect{v}_T = \frac{1}{2} \mu_H \left[\vect{E}_T(t) +
  \vect{E}_T\left(t + \Delta t\right) \right].
\end{equation}
Assuming a linear dependence of the tip velocity with the electric field is a strong simplification that nevertheless can be easily removed to incorporate more realistic dependences.  In \ref{sec:emin} we study one of them, where we impose a minimum electric field for streamer propagation.

\mysubsubsection{Branching. ---}
    We model streamer branching only phenomenologically.
    We assume that branching is a Poisson process
    characterized by the length $\ell_{\branch}$ along the streamer channel.
    Hence the probability that the streamer
    tip at $T$ branches during a time step $\Delta t$ is
    \begin{equation}
      p = v_T\; \Delta t/\ell_\branch
    \end{equation}
    We always ensure that the time step $\Delta t$ is such that $p \ll 1$.

    Once the algorithm has decided that a tip branches, the location
    of its two descendant nodes is calculated as shown in
    figure \ref{chain}d;  the locations of the
    two new tips $\vect{r}_{T'\pm}$ are symmetrical with respect
    to the location of the straight path \parref{stepping}:
    \begin{equation}
      \vect{r}_{T'\pm} = \vect{r}_T + \vect{v}_T \Delta t \pm \vect{\delta r},
      \label{branching}
    \end{equation}
    where $\vect{\delta r}$ is a random vector in the plane
    perpendicular to $\vect{v}_T$ with a bi-dimensional
    gaussian distribution with standard deviation $\ell_{\rm sib}$.


\subsection{Model parameters, specifically for positive streamers in ambient air}

\label{sect:parameters}
\begin{table*}
\begin{tabular}{lll}
Parameter and symbol & Value for positive streamers in STP air \\ \hline
Channel radius $R$ & $1 \units{mm}$ \\
Head mobility $\mu_H$ & $ 900 \units{cm^2V^{-1}s^{-1}}$\\
Line conductivity $\sigma$ & $ 9.6\dexp{-7} \units{cm\,\Omega^{-1}}$ \\
Branching ratio $\ell_\branch/R$ & $ 10$ \\
Initial separation between sibling branches $\ell_{\rm sib}/R$ & $ 0.1$ \\
\end{tabular}
\caption{\label{parameters} Parameters of our simplest model and estimated
  values for positive streamers in air at standard temperature and pressure.}
\end{table*}

Our model contains five dimensional parameters, the radius $R$
of the discharge channel,
the mobility $\mu_H$ of the channel head, the line conductivity $\sigma$,
the average channel length $\ell_{\rm branch}$ between two branching points,
and the initial separation $\ell_{\rm sib}$ between two new branches.
These parameters have to be chosen appropriately
for the system under consideration, like streamers or leaders in different gases and
at different pressures and temperatures.

For positive streamers in air at standard temperature and pressure we now estimate
their values from phenomenological observations. These values are listed in
Table~\ref{parameters}.

\begin{description}
\item{\textbf{Streamer radius $R$.}} Depending on the applied voltage,
visible streamer diameters in air at standard temperature and pressure vary
between a minimum of $\approx 0.12$ millimeter \cite{Nijdam2010/JPhD} and
3 millimeters in the experiments of Briels {\it et al.}\ \cite{Briels2008/JPhD/1}
for sharply pulsed voltages of up to 100 kV, and increase up to the order of 1 cm
in the experiments of Kochkin {\it et al.} \cite{Kochkin2012/JPhD} with a Marx-generator
delivering MV-pulses. Due to the projection
of the radiation into the 2D image plane and the nonhomogeneous excitation of emitting species in the streamer head, the radiative or visible diameter is about half of the electrodynamic diameter
that parameterizes the extension of the space charge layer around the streamer tip, i.e.
the visible diameter approximates the electrodynamic radius.  Numerical simulations
\cite{Luque2008/JPhD,Pancheshnyi2005/PhRvE} show radii in the range of 0.1 to 1 mm,
similarly to the measurements of \cite{Briels2008/JPhD/1}. As streamers of minimal
diameter generically do not branch, we have here chosen an electrodynamic radius of
$R \approx 1\units{mm}$.

\item{\textbf{Head mobility $\mu_H$.}} It was found in experiments
\cite{Briels2008/JPhD/1}
as well as in simulations \cite{Luque2008/JPhD} that the velocity of a positive streamer
strongly depends on its radius. The analysis of Naidis
\cite{Naidis2009/PhRvE} showed that the velocity of a uniformly translating streamer
also depends on the peak electric field. This is because the peak field together
with the radius determine the size of the region around the streamer head where
the electric field is above the breakdown value and where the ionization grows.
Naidis' numerical data for a fixed radiative diameter of 1 mm suggest a roughly
linear approximation $v\approx \mu_H E_p$, $\mu_H\approx 900 \,
\units{cm^2\,V^{-1}\,s^{-1}}$,
where $E_p$ is the peak electric field at the streamer ionization front.

\item{\textbf{Line conductivity $\sigma$.}} The electrical conductivity
inside a streamer channel
is dominated by the free electrons.  Most numerical simulations
\cite{Pancheshnyi2005/PhRvE,Dhali1987/JAP,Kulikovsky1997/JPhD,
Liu2004/JGRA/1,Bourdon2007/PSST,Wormeester2010/JPhD}
agree on a value of about $n_0\approx 10^{14}\units{cm^{-3}}$
electrons on the streamer axis, and a further analysis of the relation
between peak field $E_p$ and ionization density $n_0$ behind the front
can be found in~\cite{Li2007/JAP}.  If we assume a quadratic decay of the
density away from the axis up to a radius $R$, we obtain
\begin{equation}
  \sigma = 2 \pi \mathrm{e} \mu n_0 \int_0^R r \left(1 -
    \frac{r^2}{R^2}\right) \, dr
    = \frac{\pi}{2}\mathrm{e} \mu n_0 R^2,
    \label{eta}
\end{equation}
where $\mathrm{e}$ is the elementary charge, and
$\mu \approx 380 \,\units{cm^2\,V^{-1}\,s^{-1}}$ is the electron mobility \cite{Davies1971/PIEE}.
The expression \parref{eta}
yields $\sigma = 9.6\dexp{-7} \units{cm\,\Omega^{-1}}$.

\item{\textbf{Branching ratio $\ell_\branch/R$.}}
Briels \textit{et al.}\ \cite{Briels2008/JPhD} measured an
approximately linear relationship between average
branching distance and streamer radius for positive streamers in air.
We use their value $\ell_\branch/R \approx 10$,
where $R$ is the electrodynamic streamer radius.

\item{\textbf{Initial separation $\ell_{\rm sib}$ between sibling branches.}}
Finally, we used the arbitrary value $0.1 R$ for $\ell_{\rm sib}$.
The only constraints on this value are that is is much smaller than
$\ell_\branch$ and that it is of the order of $v\Delta t$, where $v$ is a typical
streamer velocity.  Below, we will find that the effect of the value of
$\ell_{\rm sib}$ on the simulations is quite weak.
\end{description}


\section{Results of the simulations}
\label{sec:results}

\begin{figure*}
\includegraphics[width=\linewidth]{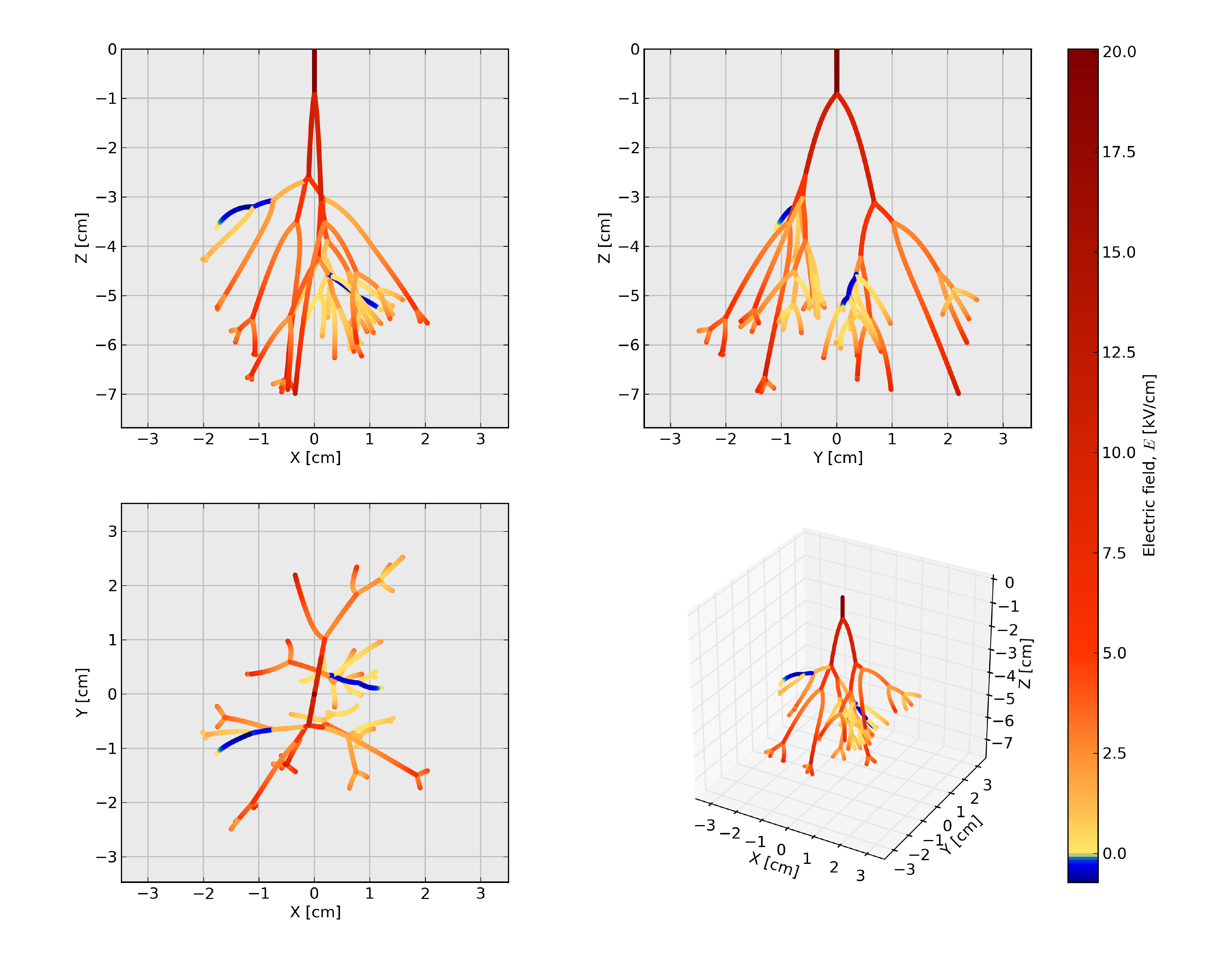}
\caption{\label{canonical}  Simulation of a positive streamer
  tree in air under normal conditions in an applied field of 15~kV/cm with the parameters of Table~\ref{parameters}.  We show the
  projections of the streamer tree on the $xz$, $yz$ and $xy$ planes
  as well as a 3d plot.  The snapshot corresponds to
  $t=80 \units{ns}$ of simulated time; at this point there are
  45 streamer branches.
  The colors of the streamer channels indicate the internal electric field, as
  described in the text.}
\end{figure*}
\subsection{Internal electric fields}
\mysubsubsection{Simulation and overall structure. ---}
Figure~\ref{canonical} shows a streamer tree simulated with the
parameters of Table~\ref{parameters} and an external
electric field $E_0=15\units{kV/cm}$ pointing downwards.
This field corresponds to about half of the classical breakdown field. We colored
the edge between two connected nodes $i$ and $j$ according to the
mean electric field in the link, defined as
\begin{equation}
  E_{ij} = \frac{\phi_i - \phi_j}{\ell_{ij}}.
\end{equation}
We chose the order of the labels $i$ and $j$ such that the electric field is
positive in the direction of streamer propagation.

The $xz$ projection of the streamer tree in Figure~\ref{canonical} (upper left) has an approximately diamond shape;
in the upper part the tree becomes wider at lower
altitude due to the repulsion between the heads whereas in a lower
part the tree gets thinner because the branches close to the center
propagate faster.  The diamond shape is typical in sprites
\cite{Cummer2006/GeoRL} and in laboratory streamers
\cite{Ebert2006/PSST,Kochkin2012/JPhD} captured before
they contact the lower electrode.
In needle-plane discharges, the strong divergence of the
electric field around the needle electrode produces a sharper
widening of the tree during the initial stages of evolution, hence in the upper part of the discharge.

\begin{figure}
 \includegraphics[width=0.75\columnwidth]{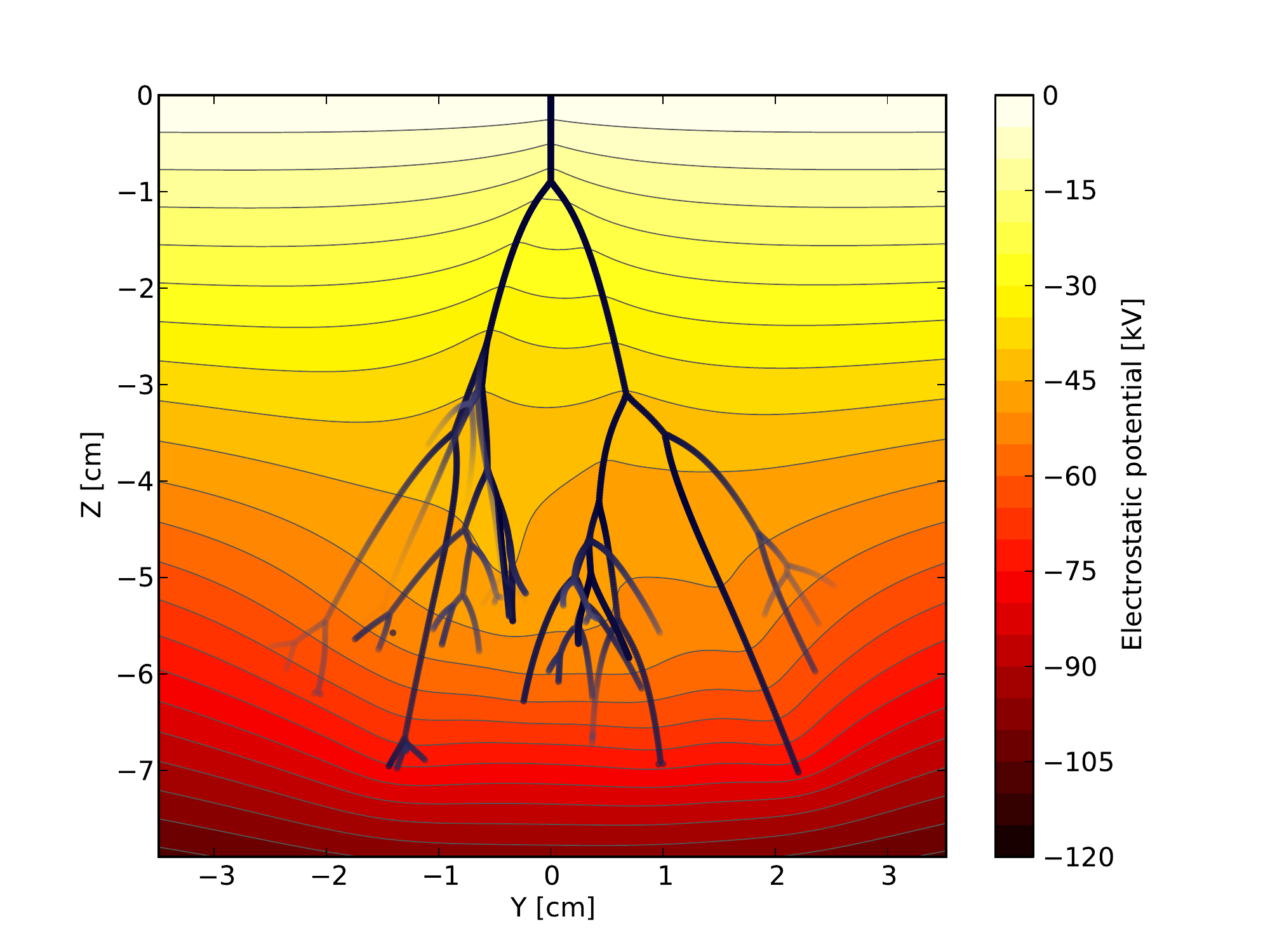}
\caption{\label{phi}  Electrostatic potential
 in the $x=0$ plane in the region surrounding the streamer tree of
 figure~\ref{canonical}.  In the projection of the streamer tree, we have
 increasingly dimmed the channels when they are further out of the $x=0$ plane.}
\end{figure}

We name the discharge structure in figure~\ref{canonical} a 
\emph{collective streamer front}; it can be interpreted as a 
``streamer of streamers.''
The many positive charges at the tips of the lower channels
have a role akin to the continuous space charge layer in a single
streamer.  Below them, they enhance the field
around the center axis; above, the field is screened.  In a single
streamer, the charge is transported to the boundary due to the
enhanced conductivity of the streamer channel; in a streamer tree,
there is a coarse-grained conductivity arising from the
many conductive filaments inside the tree.  Figure~\ref{phi} illustrates
this phenomenon by plotting the electrostatic potential in a region around
the streamer tree.  The equipotential lines are further apart inside
the tree, indicating a lower electric field, whereas they are compressed in the
volume directly in front of the tree, where the electric field is significantly
enhanced.

\mysubsubsection{Non-constant electric fields inside the streamer. ---}
The average of the internal electric fields plotted on Figure~\ref{canonical}
is close to the stability field of positive streamers
\cite{Raizer1991/book,Allen1995/JPhD,Briels2008/JPhD/1} around
$5\units{kV cm^{-1}}$.  However, we emphasize that the internal fields are not
constant, as was assumed in previous studies on streamer coronas
\cite{Pasko2000/GeoRL,Bondiou1994/JPhD,Becerra2006/JPhD/1,Arevalo2011/JPhD}.
 The field is stronger close to the streamer
head, decaying smoothly as we move upwards in the channel.  At a
branching point, the field in the parent branch exceeds that of the
two descendant branches.  This results from charge conservation: after
some transition time, the current that flows into the
branching node equals the sum of the currents flowing out; since the
currents are proportional to the internal fields, the fields in the
descendant branches must be lower than in the parent branch.

\begin{figure}
\includegraphics[width=0.6\columnwidth]{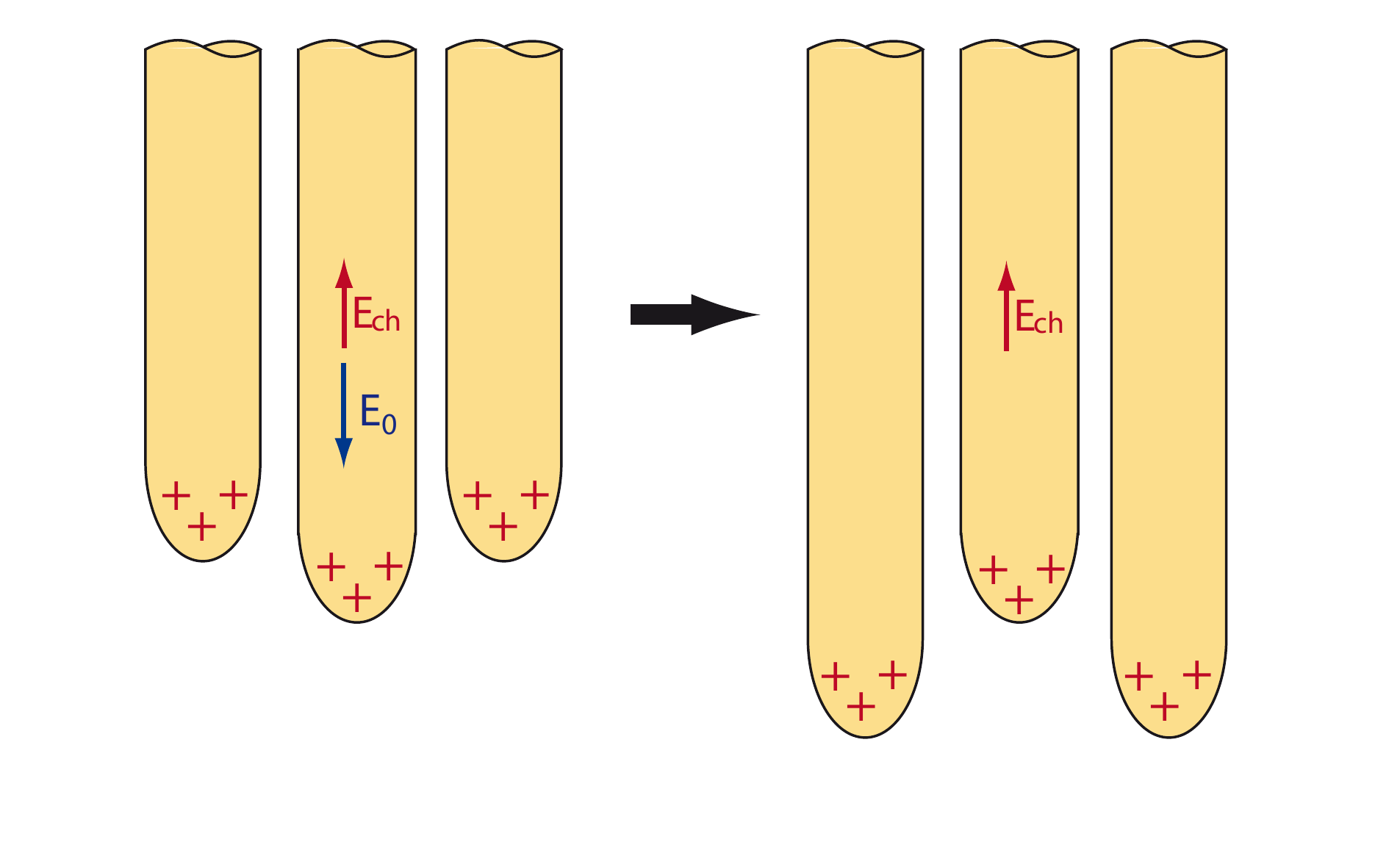}
\caption{\label{outrun}  Inversion of the inner
  electric field inside a streamer channel.  The driving electric
  field of a streamer is screened when it is overrun by neighbouring
  streamers.  In that case the streamer dies out and the charge in the
  tip is driven backwards by electrostatic repulsion. }
\end{figure}

\mysubsubsection{Field reversal. ---}
\label{sec:reversal}
A salient feature of the fields shown in Figure~\ref{canonical} is
that in some channels the fields have an opposite sign, transporting
charge backwards.  Although seemingly paradoxical, this results from some
streamers outrunning others, as outlined in Figure~\ref{outrun}.
The charges in a streamer create a field $E_{ch}$ that
oppose the external field $E_0$.  Normally $E_{ch}$ and $E_0$
add up into an internal field weaker than the external field but with
the same orientation.
Suppose, however, that the streamer is overrun by a few neighbouring
streamers carrying charges that screen $E_0$ inside
the original streamer.  Then only
$E_{ch}$ remains inside the channel, which thus starts to discharge.  In that case the streamer halts, leaving a ``dead'' channel behind.

However, our algorithm, as described in \ref{sec:growth} adds new nodes to the tree tips even for very small values of the velocity
defined in \parref{velocity}.  The resulting slow growth of these dead channels is most often irrelevant for the overall dynamics of the streamer tree but may result in unphysical behaviour, such as streamer channels slowly turning backwards.

This problem is solved by a field-velocity relation more realistic than the linear one in \parref{velocity}.  In \ref{sec:emin} we discuss the inclusion of a realistic threshold electric field for streamer propagation.

\subsection{Charge distribution in the tree}
\label{sect:charge}
\begin{figure}
\includegraphics[width=0.75\columnwidth]{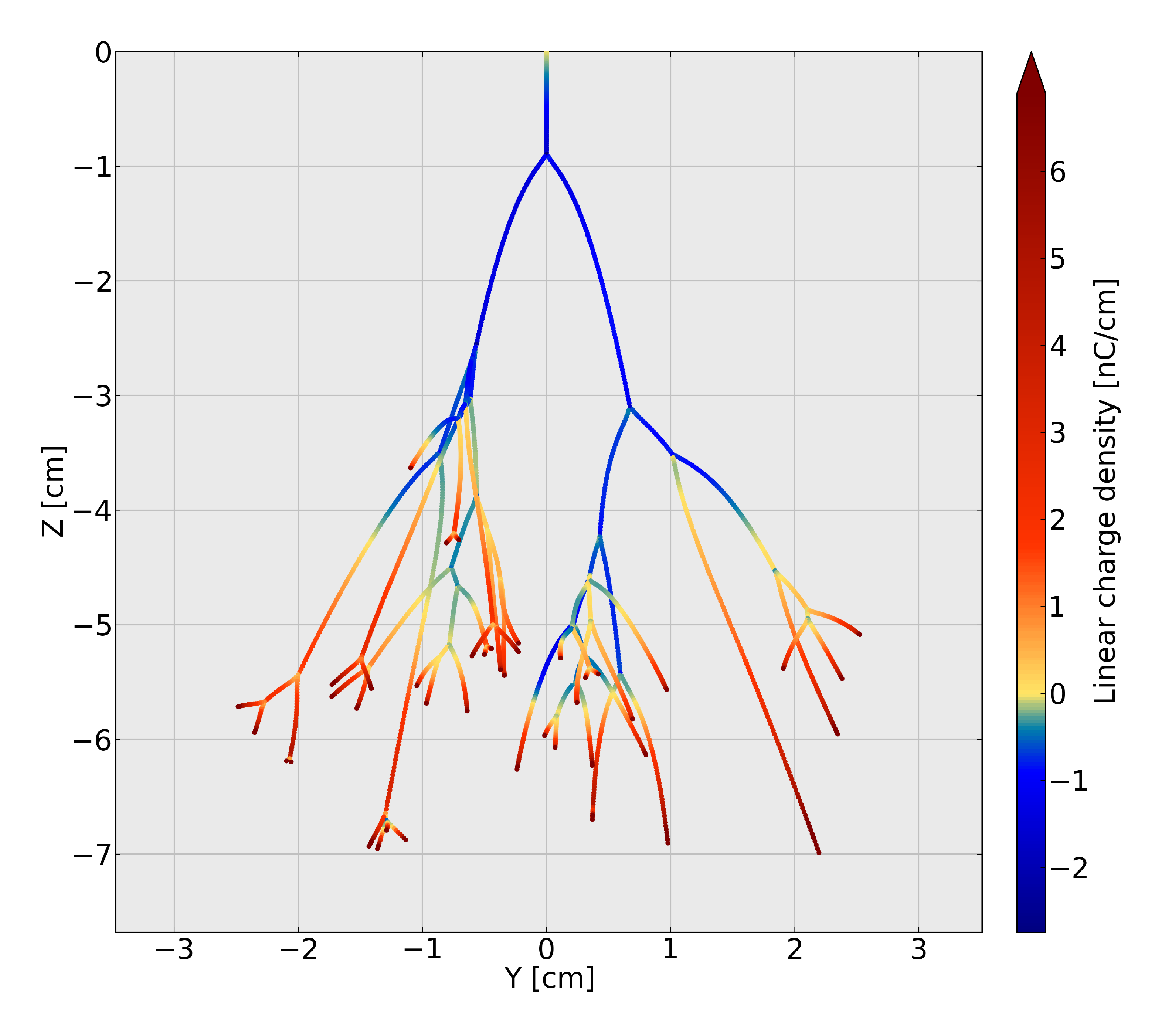}
\caption{\label{charge}  Charge distribution in the
  streamer tree of figure~\ref{canonical}.  For each node $i$ in the
  model we represent here $q_i/\ell_{\mathcal{P}(i),i}$, where $q_i$
  is the charge in the node and $\ell_{\mathcal{P}(i),i}$ is the
  length of the segment ending at $i$.  Note that the color scale is
  truncated and does not show correctly the charge density at the
  streamer tips, as they would dominate the plot.}
\end{figure}
The distribution of charges in the same
simulation as in figure~\ref{canonical} appears in figure~\ref{charge}.
To focus on the charge density inside the streamer channels, we have
truncated the color scale, which
would be otherwise dominated by the charges at the streamer heads.

Figure~\ref{charge} shows that while the lower part of the tree,
closer to the streamer tips is charged positively, the innermost
segments are negatively charged.  This resembles the negative
charging of the upper regions of sprite streamers
\cite{Luque2010/GeoRL} and arises from an analogous mechanism.  The
many channels in the external branches transport a large amount
of charge.  The fewer channels in the inner sections collect this
charge, that then gets stuck due to the lower
collective conductivity.
Hence it brings about a negatively charged inner core in the tree.

\subsection{Influence of the line conductivity}

We turn now to the influence of the line conductivity $\sigma$ of the streamer channel on the propagation and shape of the streamer tree.  We focus on this parameter because a straightforward dimensional analysis (see \ref{sec:dim}) shows that changing the line conductivity while keeping a fixed applied electric field is equivalent, after rescaling time, to a change in the external electric field with a fixed line conductivity.  Therefore the analysis described here translates directly into a study of the influence of the applied field.

\begin{figure*}
\includegraphics[width=\linewidth]{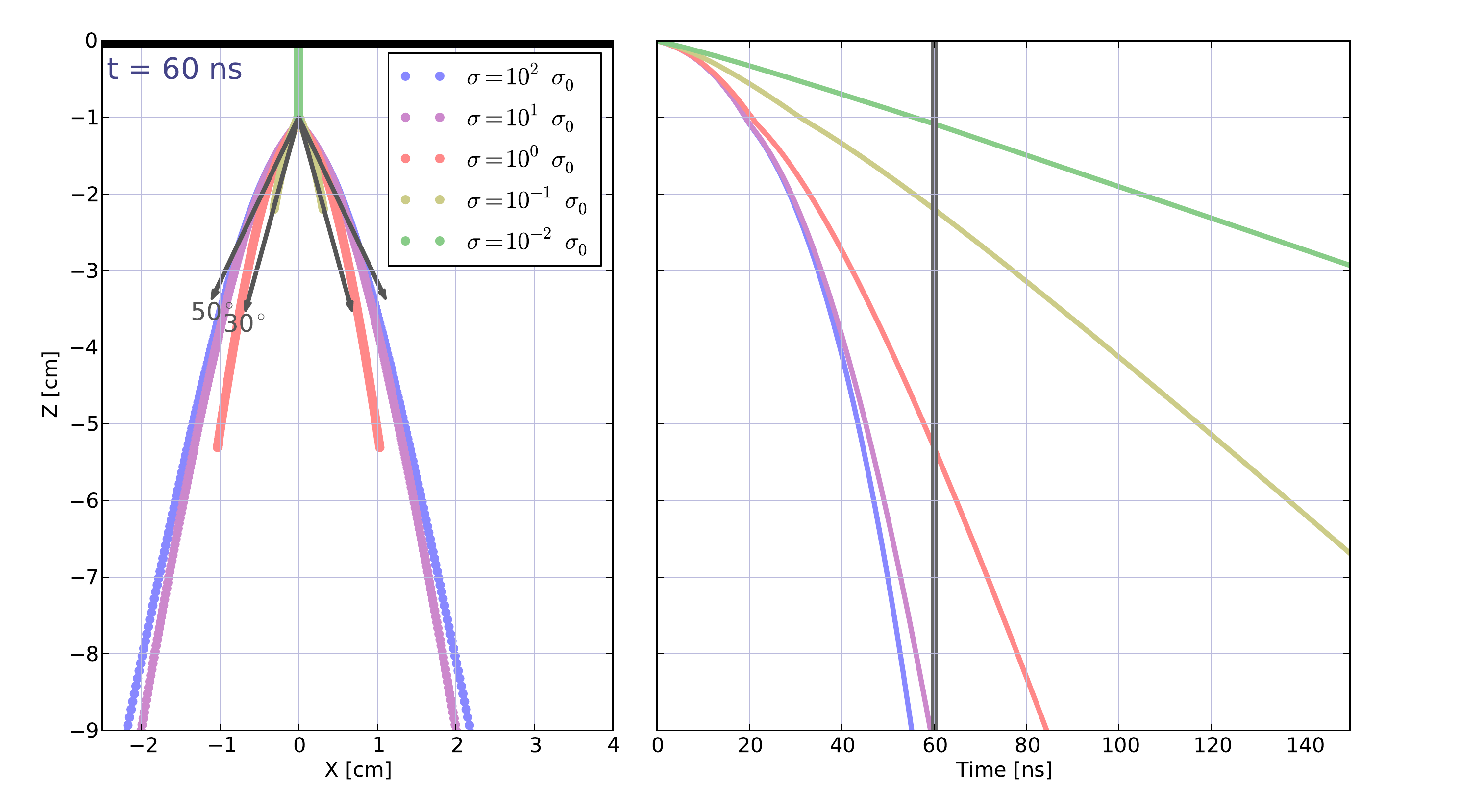}
\caption{\label{conduct_single}  Influence of the line conductivity on the propagation of a singly branched streamer.
  For different values of the line conductivity $\sigma$, the left panel
  shows a snapshot of the branch at time $t=60\units{ns}$; the right
  panel plots the location of the lowest point of the branch as a
  function of time.  The vertical line marks the time of the plots in the left panel.
}
\end{figure*}


\mysubsubsection{Branching angles. ---}
At this point, it is helpful to suppress the randomness
of the model and focus on an even simpler system.  We
run simulations where we impose a single branching point
at $z = -1\units{cm}$.  In each of these simulations,
we multiplied by a factor from $10^{-2}$ to
$10^{2}$ the line conductivity discussed above and listed in
Table~\ref{parameters}, here denoted $\sigma_0$.
Figure~\ref{conduct_single} shows the results.

The left panel of figure~\ref{conduct_single} shows the influence of
the line conductivity on branching angles.  Channels with a higher
conductivity lead to wider branching.  The reason is that charge moves
more easily along the channel and then accumulates faster
 at the streamer tips.  The electrostatic repulsion between
both heads is thus stronger and they diverge more sharply.

However, figure~\ref{conduct_single} shows that this mechanism is quite weak.  Although it is theoretically possible to infer the channel conductivities from branching angle measurements, such as those by Nijdam \emph{et al.} \cite{Nijdam2008/ApPhL}, the
dependence seems too weak to be useful, given the natural variation and the measurement uncertainties of branching angles.  In
figure~\ref{conduct_single} we mark with arrows the
branch-to-branch angles $30^\circ$ and $50^\circ$ from the branching point to underline that all conductivities agree with the branching angles of $(39.7\pm13.2)^\circ$ reported in reference~\cite{Nijdam2008/ApPhL} for positive streamers in air at atmospheric pressure.

\mysubsubsection{Velocity. ---}  On the right panel of
figure~\ref{conduct_single} we plot the propagation distance of the
streamers as a function of time for the same simulations as in the previous
section.  We see a significant speed-up of the propagation with
increasing channel conductivity.  Again, the
increased charge transport and accumulation at the streamer
tip explain this behaviour.

Another feature of Figure~\ref{conduct_single} is that the streamers with line conductivity $10 \sigma_0$ and $10^2 \sigma_0$ propagate almost at the same speed despite an order of magnitude difference in $\sigma$.  The reason is that they approach the high-conductivity regime, where the charge distribution in the streamer adjusts instantaneously to changes in the streamer length.  The reference value $\sigma_0$ is about a factor 10 below this limit, implying that the finite streamer conductivity is still relevant for the streamer propagation.

\subsection{Influence of $\ell_{sib}$}
\begin{figure}
\includegraphics[width=0.75\columnwidth]{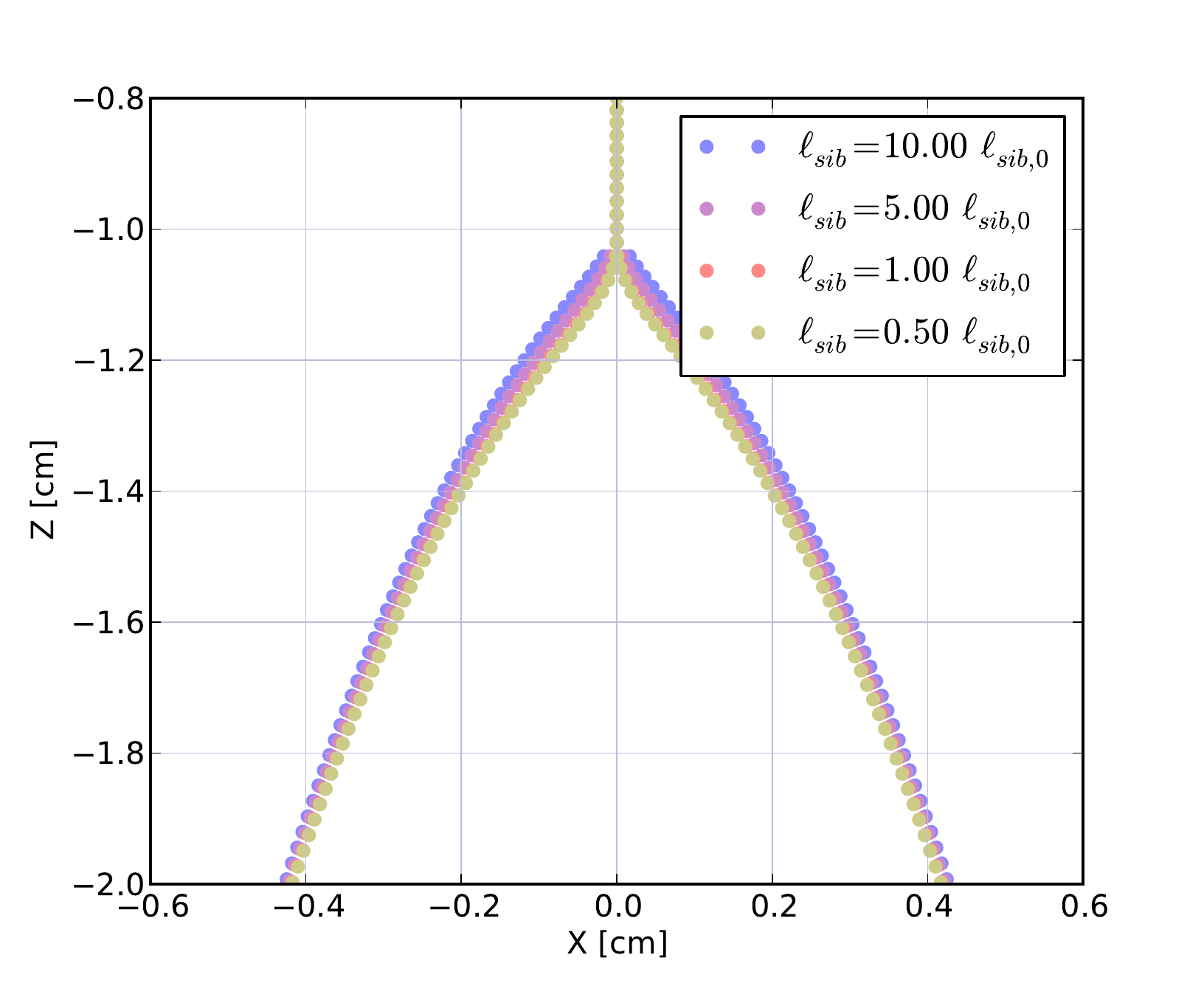}
\caption{\label{sigma}  Four simulations with
  different values of $\ell_{sib}$.  In the figure legend, the
  $\ell_{sib,0}$ refers to the value in table
  \ref{parameters}, $\ell_{sib,0} = 0.1 \units{mm}$.}
\end{figure}
As we mentioned above, $\ell_{sib}$ does not substantially
influence the simulations as long as it stays within reasonable
physical bounds.  To investigate this, we run simulations where we
changed $\ell_{sib}$ from one tenth to twice the value in table
\ref{parameters}.  As in the previous section, in these simulations
we forced the streamers to branch uniquely at a prescribed location
$z=-1\units{cm}$.  The outcome appears in figure \ref{sigma}.

Simulations with very different $\ell_{sib}$ behave similarly.
After a short transient, the electrostatic repulsion between the two
sibling branches strongly dominates their propagation.  About $1
\units{cm}$ below the branching point, the trajectories of
simulations with different $\ell_{sib}$ are barely separated.  We conclude
that $\ell_{sib}$ which was introduced as a numerical parameter, does not 
influence he results much.

\subsection{Reconnection}
\begin{figure*}
\includegraphics[width=\linewidth]{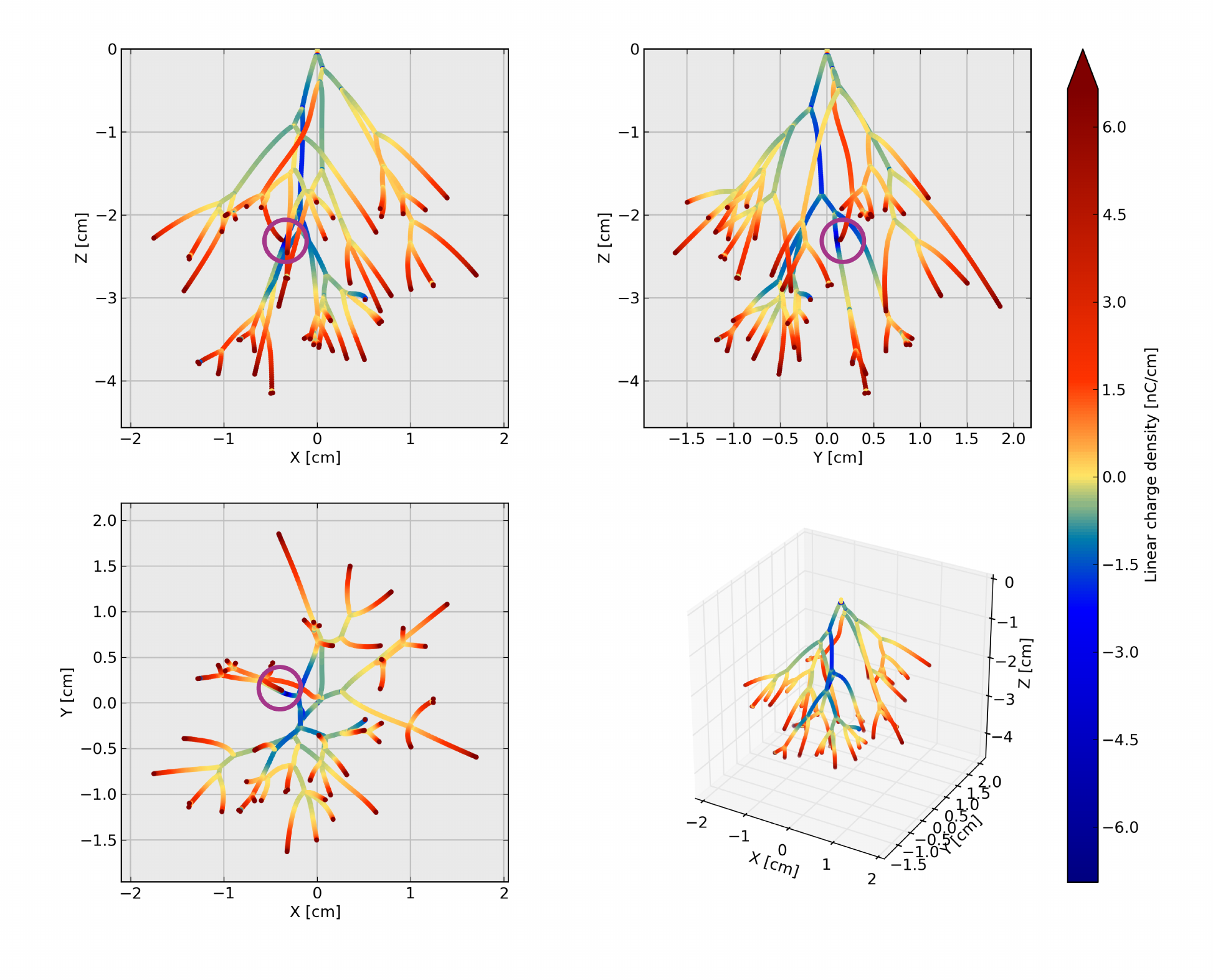}
\caption{\label{reconnection}  A reconnection event.  The
  parameters of this simulation are those listed in
  Table~\ref{parameters}.  We
  show here a snapshot of the charge distribution at time
  $t=160.75 \units{ns}$.  The circles mark
  the place of reconnection in the three projections; it is clearly
  seen in the $xz$ projection.}
\end{figure*}

 Let us now use our model to investigate the reconnection of streamer
channels inside a tree.  In a reconnection event, a streamer head is
attracted towards a pre-existing
channel.  This should not be confused with streamer
\emph{merging}, where two streamer heads expand to form a single channel
\cite{Luque2008/PhRvL,Bonaventura2012/PSST}.

Streamer reconnection has been observed
both in laboratory discharges \cite{van_Veldhuizen2002/JPhD,Nijdam2009/JPhD}
and in high-speed sprite observations
\cite{Cummer2006/GeoRL,Stenbaek-Nielsen2008/JPhD,Montanya2010/JGRD}.
Nijdam \textit{et al.} \cite{Nijdam2009/JPhD}, reviewed the recorded examples
of reconnection and extended them with new experimental data.
Using stereoscopy, they were able to discriminate between actual
reconnection and ambiguous observations resulting from projecting the
3d streamers into the camera plane.  They concluded that 
reconnection of positive streamers in laboratory experiments is indeed
frequent but consists in a thinner, slower streamer moving towards the channel of a thicker, faster streamer that had already contacted the cathode.  After this contact,
the ionized streamer channel charges negatively and attracts
the streamer heads surrounding it, still positively charged.
Although commonplace in the laboratory, this mechanism does not
explain the observations of streamer reconnection in sprites, where a
lower electrode does not exist.
Here we will limit ourselves to the study of this latter kind of
reconnection, where a lower electrode does not exist or is not essential.
We henceforward restrict the meaning of \emph{reconnection} to this type of
event only.  In this restricted sense, reconnection has not been
unambiguously observed in laboratory experiments.

We frequently observe reconnection events in our model.
Figure~\ref{reconnection} shows an example; there, a lagging streamer
is attracted to the stem of a sub-tree that has propagated much farther.
This pattern is generic to all the reconnection events that
we found in our simulations.  The picture shows that the reason is that,
as explained in section~\ref{sect:charge}, the inner branches of the tree
acquire a negative charge; usually, most of the channels in that volume are
similarly negatively charged but if a lagging streamer propagates through
the inner sections of the tree, its positive charge is attracted and
reconnects to a negative, inner branch.  To put it concisely, the extremal
branches are attracted towards the internal ones.

\begin{figure}
\includegraphics[width=1.0\columnwidth]{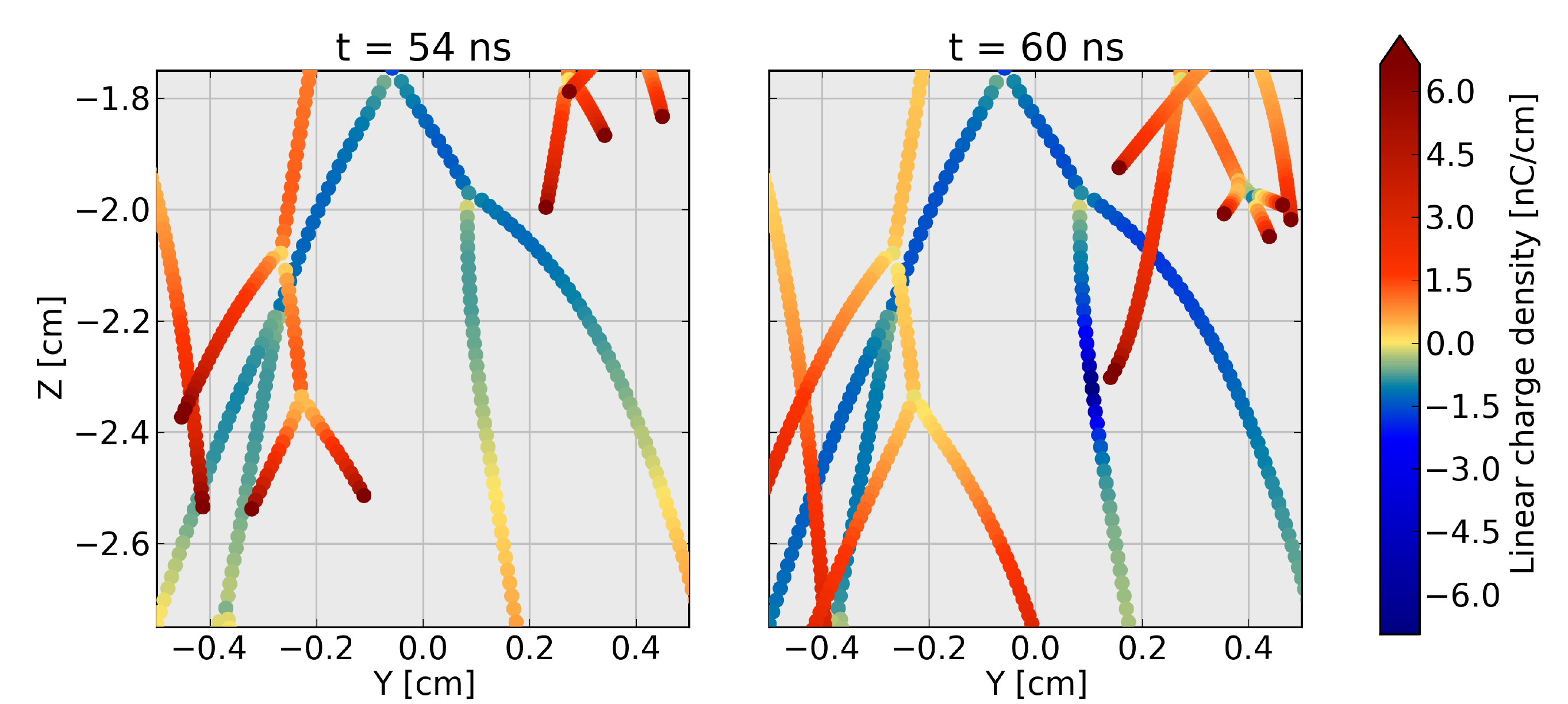}
\caption{\label{reconnection_zoom}  Zoom of the reconnection
  event of figure~\ref{reconnection} at two time steps and
  projected onto the xz plane.  A positively charged streamer
  head approaches a pre-existing, negative channel.   The negative
  charge in the channel induced by the head is clearly visible in the
  latest time step (right panel) but an earlier time step (left panel) shows
  that the channel had already a negative charge before the interaction.
  Note also that the other branches at the right of the picture also
  charge negatively, even though they are not directly involved in the
  reconnection.}
\end{figure}

In figure~\ref{reconnection_zoom} we zoom into the reconnection of
figure~\ref{reconnection} and plot two snapshots of the charge distribution.
We see that as the head approaches the channel, it induces a significant,
additional negative charge in the pre-existing channel.  The relevance of
these induced charges in a conductive channel was pointed out by
Cummer \textit{et al.} \cite{Cummer2006/GeoRL}.  Nevertheless, our simulations
suggest that the initial attraction of a head towards a channel is possible
only in cases where that channel has the opposite charge.  The induced charges
dominate only when the head is already very close to the channel.

We speculate that reconnection (in our restricted sense) has
not been observed in laboratory discharges because their innermost branches do
not charge negatively or do not do it strongly enough.  We offer two possible
reasons for this. (a) That the needle-electrode geometry most often employed in
the laboratory, by imposing higher and divergent electric fields around 
the anode, discharges the negative charges in that region faster and reduces streamer interaction. (b) That the reduced propagation
length imposed by the cathode does not allow the tree enough time to reconnect.
Most likely, there is a combination of both (a) and (b) at play; 
and finally, in the laboratory experiments \cite{Nijdam2009/JPhD}, only for
sparse trees with less than about 50 streamers the full 3D structure can
be reconstructed which gives a bias in the observations.

\begin{figure}
\includegraphics[width=\columnwidth]{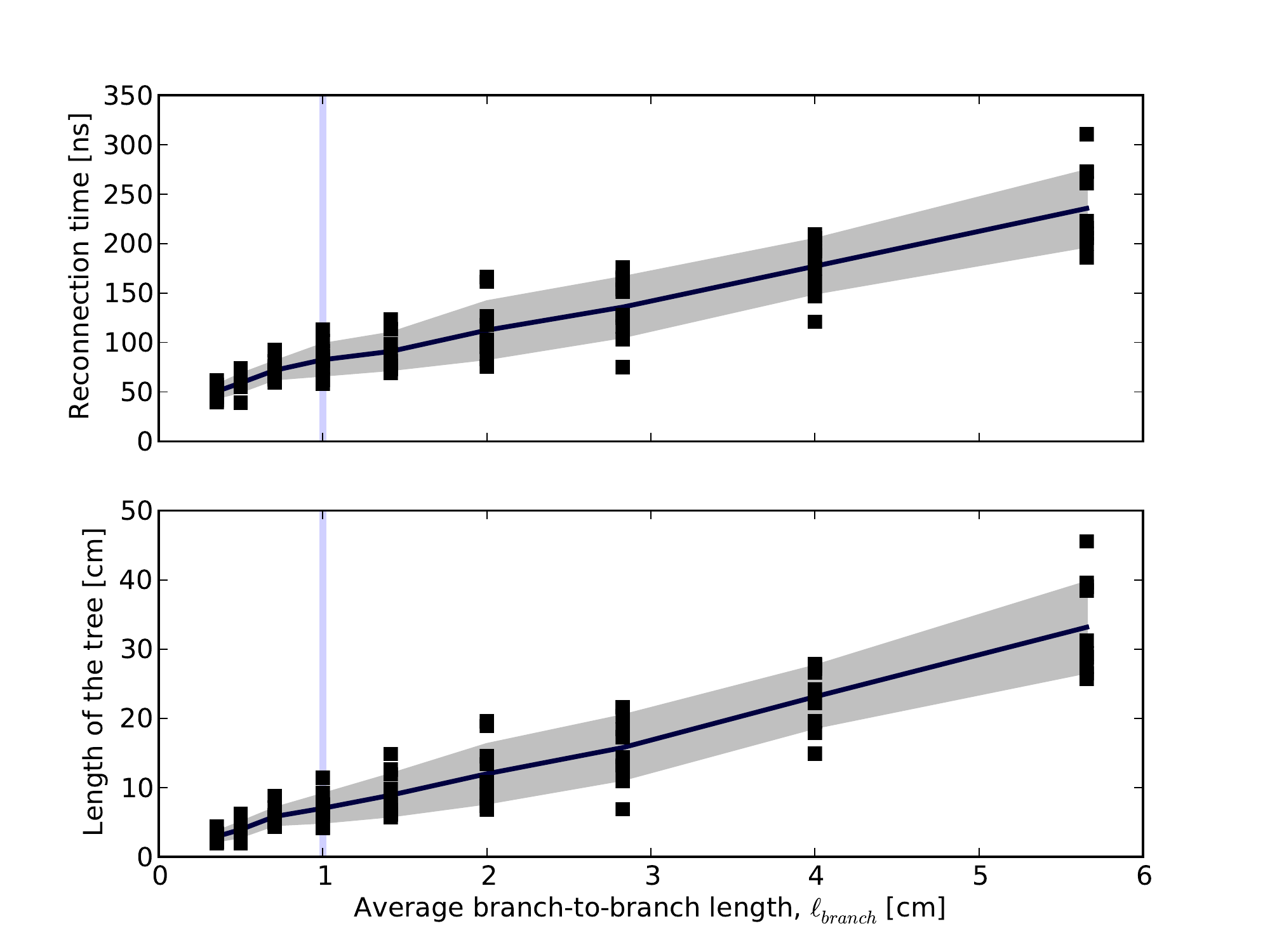}
\caption{\label{wbranch}  Dependence on the branching frequency
  $\ell_{branch}$ of the time to the first reconnection event and the total tree
  length.  Here the total tree length is the largest absolute value of the $z$
  coordinate of any point in the tree.  For each value of $\ell_{branch}$ we run
  10 simulations, plotted black squares; the continuous line represents the
  mean of these 10 simulations and the shaded area includes one standard
  deviation around the mean.  The vertical line marks the standard value
  $\ell_{branch}=1\units{cm}$ from table~\ref{canonical}.}
\end{figure}

To investigate further whether we should expect to see streamer reconnection in
laboratory experiments, we can tune the parameters in our model and make
reconnection more or less likely.  In particular, we may force the streamers
to branch more or less frequently by varying the parameter $\ell_{branch}$.  We
used values from $0.35\units{cm}$ to $5.5\units{cm^{-1}}$ and for each
value we run 10 simulations up to the time of the first reconnection.  The
results are plotted in figure~\ref{wbranch}.

For the standard value $\ell_{branch}=2\units{cm}$ the plot indicates that
we need a gap of about $7\units{cm}$ between electrodes to have a significant
chance of observing reconnections; if $\ell_{branch}$ would increase to $2.85\units{cm}$, one would need a gap of more than $12\units{cm}$.  Given
the uncertainties and approximations in our model and point (a) discussed above
we believe that laboratory discharges would also reconnect if they are given
enough space.

\section{Summary and conclusions}
\label{sec:summary}

Discharge tree models constitute the highest level in space in the hierarchy of electrical discharge models.  While in the past they were frequently based on phenomenological assumptions, we here present a model that rests on results and insights from fluid models, which in turn depend on the micro-physics of collisions described by particle or Boltzmann-equation models.  As P.W. Anderson famously remarked \cite{Anderson1972/Sci}, each new level in such a hierarchy usually contains nontrivial, sometimes surprising, physics that are not immediately apparent from our understanding of the lower levels.

Here we have shown that even the simplest tree model with self-consistent charge transport leads to new insights into the distribution of charges and electric fields and into the process of streamer reconnection.  Our model also reveals the qualitative self-similar nature of collective streamer fronts, where the full structure can be seen as a ``streamer of streamers'', i.e., a scaled-up analogue of each of the streamers that compose it.

Clearly many elements of streamer physics have not been incorporated here into our model.  A non-exhaustive list includes the dynamical selection of streamer diameters, the different ionization levels created in the streamer head depending on the field enhancement, and the changes in the channel conductivity due to attachment processes, the extension to negative streamers and to the gradient in air density experienced by sprite streamers in the upper atmosphere.  Forthcoming investigations shall address these issues.

\begin{acknowledgments}
This work was supported by the Spanish Ministry of Science and
Innovation, MICINN under project AYA2011-29936-C05-02 and and by the Junta de Andalucia, Proyecto de Excelencia FQM-5965.  AL acknowledges support by a Ram{\'o}n y Cajal contract, code RYC-2011-07801.  UE acknowledges support from the European Science Foundation (ESF) for a short visit within the ESF activity entitled 'Thunderstorm effects on the atmosphere-ionosphere system' (TEA-IS).
\end{acknowledgments}

\appendix

\section{Notes on the numerical implementation}
\label{AppNumerics}

\subsection{Convergence of numerical time stepping}
\begin{figure}
\includegraphics[width=\columnwidth]{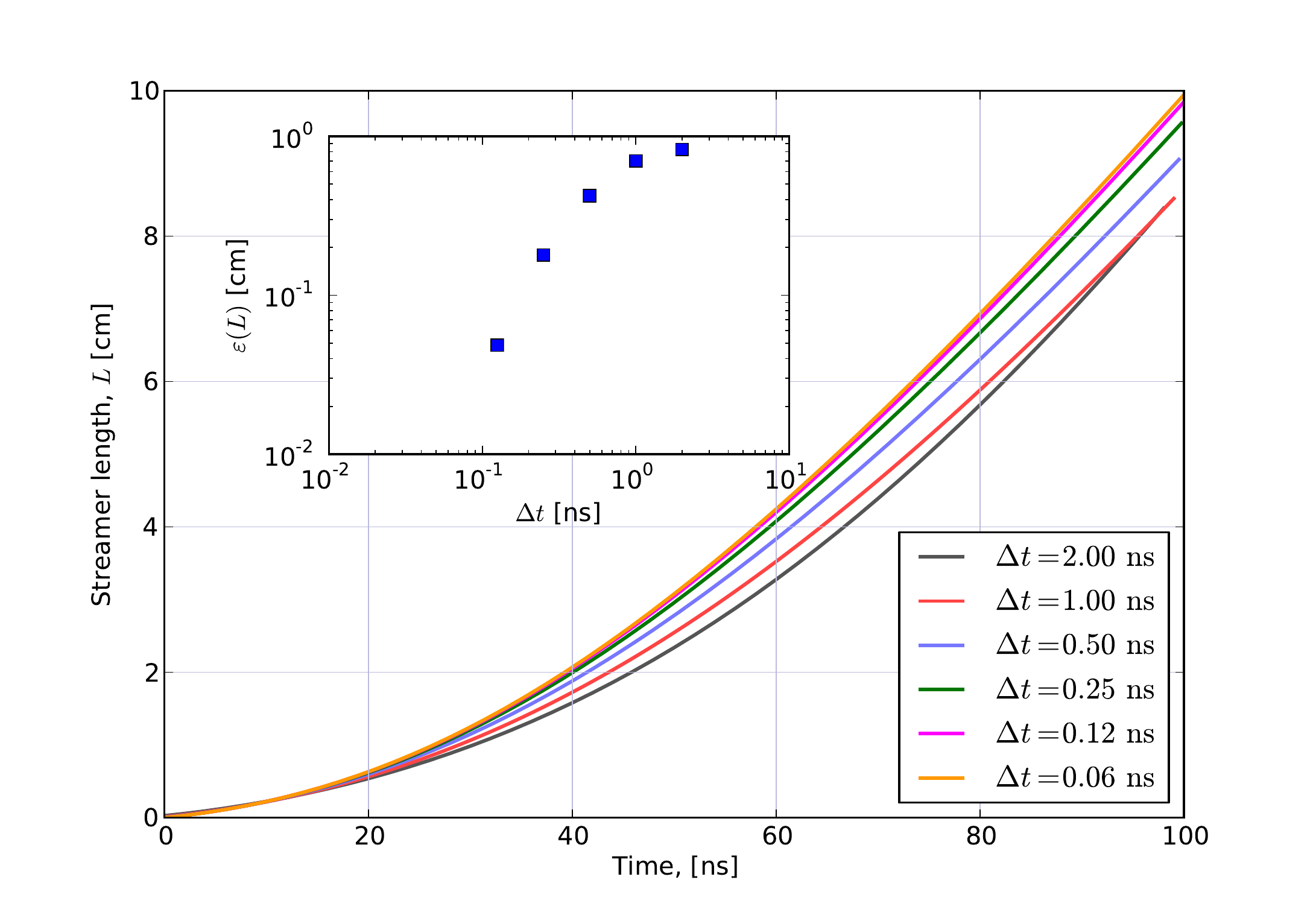}
\caption{\label{convergence}  Convergence of the model
  simulations with decreasing time step $\Delta t$.  The main figure shows
  the evolution of the streamer length
  with $E_0 = 15\units{kV/cm}$ for different time steps. The inset
  plots the estimated error of each simulation as a function of
  $\Delta t$.  Here $\varepsilon(L)$ is the root mean square of the
  difference between streamer lengths of a simulation and the most
  accurate simulation, $\Delta t = 0.06 \units{ns}$.}
\end{figure}
A necessary condition for the numerical calculation of the model is that it
converges for decreasing time step $\Delta t$.
To check this, we run deterministic simulations (with $\ell_\branch=0$) with an
external electric field $E_0 = 15\units{kV/cm}$ and various $\Delta t$.
Figure~\ref{convergence} shows the length of the streamer channel as a
function of time; the simulations converge to a solution once the
time steps are shorter than about $0.25 \units{ns}$.

Therefore in all simulations in this paper we use $\Delta t
= 0.25 \units{ns}$.

\subsection{Numerical solution of the electrostatic problem}
We are calculating all interactions between pairs
of charged nodes and therefore our computation time scales as
$O(N^2)$.  This is the main limitation in the size of trees that we can
efficiently simulate.  To overcome this limitation we also
implemented the Fast Multipolar Method (FMM) which is able to solve
the electrostatic problem with $O(N)$ computations up to an
arbitrarily good approximation.  However, the kernel
in \parref{potentials} is not the Poisson kernel for $R\neq 0$ and
although we restricted the FMM only for distant interactions with
$r_{ij} \gg R$, we run into problems around the cutoff.  Besides, we
found that due to the overhead of the FMM, it was advantageous only
for $N$ larger than a few thousand and all the simulations reported
here are below that threshold.  Each of the simulations that we show took a
few hours in a modern desktop computer.

\section{An improved model for the propagation of streamer tips}
\label{sec:emin}

For the sake of simplicity we have assumed a linear dependence of the velocity with the electric field at the streamer tips.  As we discussed in section \ref{sec:reversal}, often this leads to slow streamers that keep propagating even when the surrounding electric field is very small.  This contradicts both experimental observations and our theoretical understanding, where impact ionization is essential for streamer propagation.  A more realistic model must include a minimum field for streamer propagation.

Taking an electrodynamic streamer radius $R=1\units{mm}$ ($\sim$ radiation diameter), the analytical calculations in \cite{Naidis2009/PhRvE} are well fitted by
\begin{equation}
  \label{naidis}
  v_T = \mu_H \max\left(0, E_T - E_{min}\right),
\end{equation}
where the head mobility is now $\mu_H = 3200 \units{cm^2V^{-1}s^{-1}}$ and the threshold field for propagation is
$E_{min} = 100\units{kV/cm}$.

However, \parref{naidis} presents a new problem in our plane-electrode geometry.  If the applied field $E_0$ is lower than $E_{min}$, the tree will not start to propagate by itself.  The natural solution is to implement a needle-plane geometry; here we simulated a 1 cm-needle by starting the tree from a vertical chain of 10 nodes separated by 1 mm.  With $E_0=15\units{kV/cm}$ this was enough to initiate a tree.

\begin{figure*}
\includegraphics[width=\linewidth]{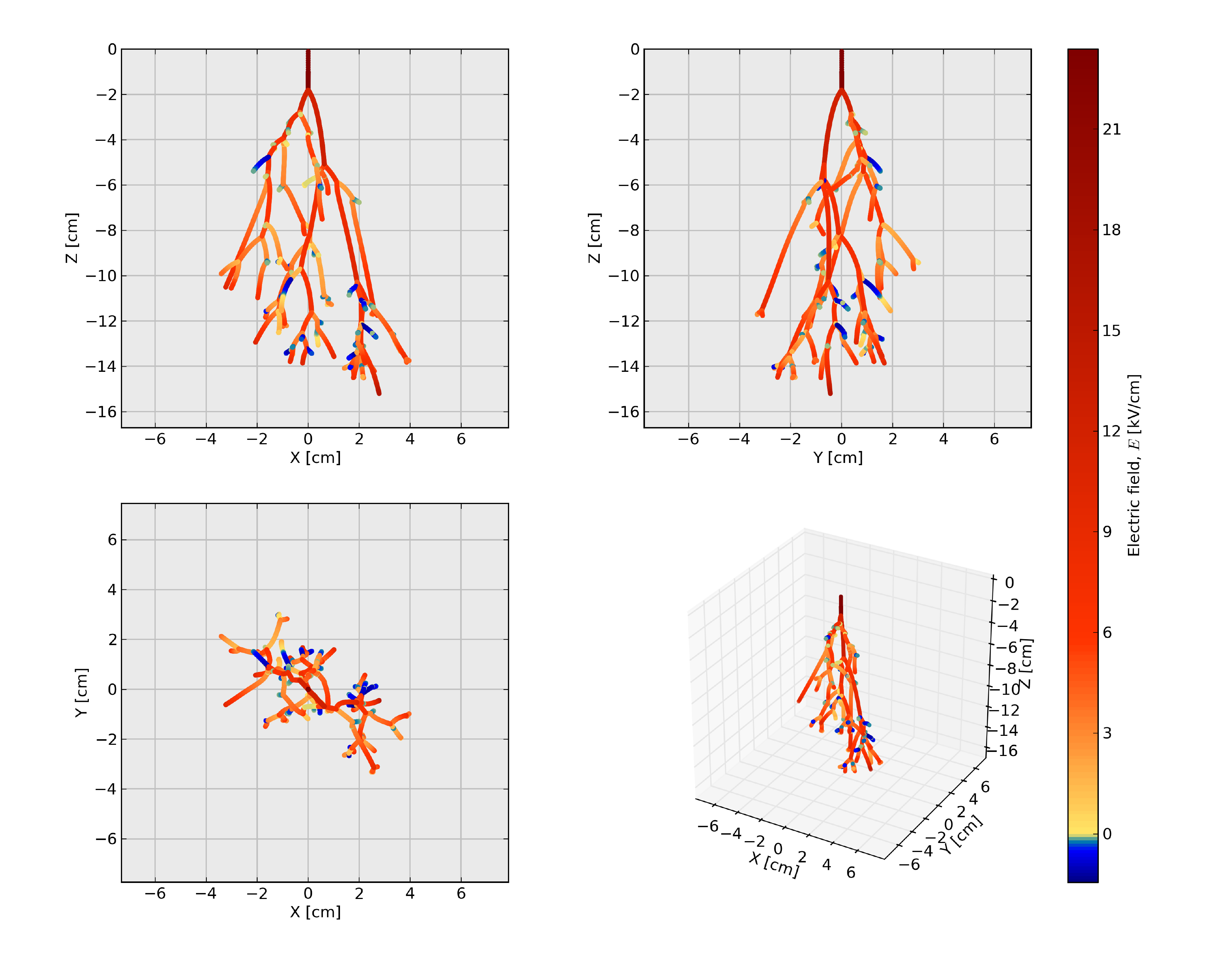}
\caption{\label{emin}  Streamer
tree with tips growing according to equation \parref{naidis} in the     text; all other parameters are the same as in figure~\ref{canonical},
listed in table~\ref{parameters}.  Here we show a snapshot of the internal electric fields at time $t=125\units{ns}$.}
\end{figure*}

In figure \ref{emin} we show the tree created in a simulation where head velocities are as in \parref{naidis}.  All other parameters are the same as in figure \ref{canonical} in the main text.  The most remarkable feature in the tree of figure \ref{emin} is the multitude of short channels that punctuate the trails of longer streamers.  Often, these channels are so short that they are seen only as a sudden change in the direction of the branch.  Both short branches and apparent changes in streamer direction are observed in laboratory photographs of streamer trees; they are very common in nitrogen discharges but they also appear in air (see e.g. figure 1 in reference \cite{Briels2008/ITPS/1}).

\section{Dimensional analysis of the model}
\label{sec:dim}
The dimensional quantities of our model are those listed in
Table~\ref{parameters} plus the vacuum permittivity
$\epsilon_0 = 8.85\dexp{-14}\units{CV^{-1}cm^{-1}}$.  Straightforward
dimensional analysis leads to the characteristic scales listed on
table~\ref{characteristic}.  Note that the characteristic scales
follow the Townsend scaling laws \cite{Ebert2010/JGRA/c}; our results
can be rescaled to any gas density.

A remarkable feature of table~\ref{characteristic} is the high value
of the characteristic electric field,
$\mathcal{E}=2260\units{kV/cm}$.  This value is much higher than what
is commonly observed in atmospheric pressure streamers and also in our
simulations.  The reason is that $\mathcal{E}$ defines the electric
field created by a typical electron density confined in a typical
streamer volume.  However, $\mathcal{E}$ does not take into account
that most of the electron density is screened by a similar density of
positive ions.  The weak-field limit in our model, where all electric
fields are much lower that $\mathcal{E}$, is therefore equivalent to quasi-neutrality; namely that the electron and ion densities $n_e$, $n_\pm$ satisfy $|n_+ - n_- - n_e| \ll n_e$.

One can use the values in Table~\ref{characteristic} to
derive a dimensionless model where the only parameters are
$R/\ell_\branch \approx 1/20$ \cite{Briels2008/JPhD} and, for a given
external electric field $E_0$, the ratio $E_0/\mathcal{E}$.  An immediate consequence is that these two dimensionless quantities
fully determine the geometric properties of a streamer tree, such as angles and length ratios.

\begin{table}
\begin{tabular}{lll}
Magnitude & Characteristic scale & Value at atmospheric pressure \\ \hline
Length & $R$ & $1\units{mm} $ \\
Electric field & $\mathcal{E}=\sigma/4\pi\epsilon_0 \mu_H R$ & $2260\units{kV/cm}$ \\
Velocity & $v = \mu_H \mathcal{E}$ & $2\cdot10^7\units{m/s}$ \\
Time & $\tau = R/v$ & $0.12\units{ns}$
\end{tabular}
\caption{\label{characteristic} Characteristic scales of the streamer
  tree model.}
\end{table}


\bibliography{library}

\end{document}